HOW TO PRINT THIS PAPER:
1. Split this file into a TeX code file + 3 Postscript files. The separation
between the files is indicated by the keyword "FIGURE" in capitals.
2. Latex the code file (RevTeX 3.0). The output is in landscape format.
When using the "dvips" command on a UNIX station, the option "-t landscape"
should be used, as in "dvips -t landscape paper.tex".
3. Each Postcript file should start with "
be deleted. Then the files may be printed by the UNIX command "lpr".
DELETE UP TO NEXT LINE

\documentstyle[twocolumn,prb,aps]{revtex}
\begin{document}
\textwidth=24cm
\textheight=18cm
\title{Haldane gap in the quasi one-dimensional nonlinear $\sigma$-model}
\author{David S\'en\'echal}
\address{ Centre de Recherche en Physique du Solide et D\'epartement de
Physique,}
\address{Universit\'e de Sherbrooke, Sherbrooke, Qu\'ebec, Canada J1K 2R1.}
\date{May 1993}
\maketitle
\begin{abstract}
\centerline{ABSTRACT}
This work studies the appearance of a Haldane gap in quasi one-dimensional
antiferromagnets in the long wavelength limit, via the nonlinear
$\sigma$-model. The mapping from the three-dimensional, integer
spin Heisenberg model to the nonlinear $\sigma$-model
is explained, taking into account two antiferromagnetic couplings: one along
the chain axis ($J$) and one along the perpendicular planes ($J_\bot$) of a
cubic lattice. An implicit equation for the Haldane gap is derived, as a
function of temperature and coupling ratio $J_\bot/J$. Solutions to these
equations show the existence of a critical coupling ratio beyond which a gap
exists only above a transition temperature $T_N$.
The cut-off dependence of these results is discussed.
\end{abstract}
%
\section{Introduction}
It is by now well established\cite{Haldane,Night} that the excitation spectrum
of the spin 1 antiferromagnetic Heisenberg model has a mass gap (the Haldane
gap) in one dimension, whereas it is gapless in higher
dimensions\cite{Kennedy}. Experimentally, integer spin chains are realized in
quasi one-dimensional compounds in which the antiferromagnetic coupling $J$
along the chain direction is much higher than the transverse coupling
$J_\bot$. The Haldane gap can then be observed via neutron
scattering\cite{Buyers} or electron spin resonance\cite{ESR}.
It is the ratio $R=J_\bot/J$ which determines the degree of `quasi
one-dimensionality' of the material. For instance, $R$ is
estimated\cite{Buyers} to be $\sim0.02$ in CsNiCl$_3$, while it is certainly
much lower ($\sim 0.0006$) in NENP.\cite{Renard}
In CsNiCl$_3$ one-dimensional behavior (i.e. the existence of the gap) is
oberved above a critical temperature (the N\'eel temperature $T_N$) of about
$5K$. On the other hand, one-dimensional behavior is observed in NENP at
temperatures as low as can be reached. The suggests the existence of a critical
ratio $R_c$ below which the system is one-dimensional in character, whatever
the temperature.

This was argued in Ref.\onlinecite{Chubukov}, wherein corrections to the
spin-wave spectrum were claculated in perturbation theory for the anisotropic
Heisenberg model in two dimensions. More recently,
Azzouz and Dou\c cot\cite{Azzouz} have performed a mean-field theory analysis
on the Heisenberg model on a square lattice:
\begin{equation}
\label{Heisen}
H = J\sum_{\langle ij\rangle} {\bf S}_i\cdot{\bf S}_j +
J_\bot \sum_{\langle lm\rangle} {\bf S}_l\cdot{\bf S}_m
\end{equation}
wherein the first sum runs over nearest neighbor spins along the chains, and
the second sum runs over nearest neighbors in the direction perpendicular to
the chains. They found a critical ratio given, in the large $s$ approximation,
by $R_c \sim se^{-2\pi s}$, where $s$ is the value of the integer spin.

In this work we will perform a similar analysis, but within the
three-dimensional non-linear $\sigma$-model, and at finite temperature. The
technique used has been described in Ref.\onlinecite{Senechal},
wherein the temperature dependence of the gap was calculated for the
purely one-dimensional case.

The value we will find for $R_c$ is very sensitive to the cut-off prescription,
but if we fix the latter with the help of the numerical result
$\Delta_0\sim.41J$ for the purely one-dimensional Haldane gap at zero
temperature, we find the critical ratio $R_c\sim0.026$. However, we believe
that this estimate should be regarded with great caution. We also illustrate
the $R$ dependence of the N\'eel temperature, of the zero-temperature gap, and
the temperature dependence of the gap for various values of $R$.

\section{The anisotropic nonlinear $\sigma$-model}
The aim of this section is to show how the Heisenberg Hamiltonian
(\ref{Heisen})  may be replaced, in dimension 3 and in the continuum limit, by
the anisotropic nonlinear $\sigma$-model, with Lagrangian density
\begin{equation}
\label{nonlinearS}
{\cal L} = {1\over 2g}\left\{
{1\over v} (\partial_t{\bf m})^2 -
v (\partial_z{\bf m})^2 - Rv (\partial_x{\bf m})^2
 - Rv (\partial_y{\bf m})^2 \right\}
\end{equation}
wherein
\begin{equation}
\label{GandV}
v = 2Jas\sqrt{1+2R}\qquad
g = {2a^2\over s}\sqrt{1+2R}
\end{equation}
and where $a$ is the lattice spacing; the field ${\bf m}$ is a unit vector:
${\bf m}^2=1$; this constraint alone makes the theory non trivial.
Readers willing to accept Eqs (\ref{nonlinearS}) and (\ref{GandV}) may proceed
to the next section.

The mapping from the Heisenberg Hamiltonian to the nonlinear $\sigma$-model
has been done explicitly before in various ways\cite{Haldane,Fradkin},
but mostly in dimension 1. Here we wish to show explicitly how the mapping is
done in dimension 3, with anisotropic couplings.
We first need to write down the action for a single site, using spin coherent
states. We define a unit vector $\bf n$ such that ${\bf S}= s{\bf n}$. Then
the kinetic term is well-known to be the Wess-Zumino action:\cite{Fradkin}
\begin{equation}
\label{actionWZ}
S_{WZ} = \int_0^T dt~K =
s\int_0^1 d\tau\int_0^T dt~ {\bf n}\cdot(\partial_\tau{\bf n}
\times \partial_t{\bf n})
\end{equation}
Here the time $t$ runs from 0 to some finite period $T$ and $\tau$ is an
auxiliary coordinate introduced in order to parametrize, along with $t$, the
spherical cap delimited by the curve  ${\bf n}(t)$ from $t=0$ to $t=T$. For
more details, the reader may consult Ref.\onlinecite{Fradkin}. For our
purpose, it is sufficient to know that the variation of the action upon a small
change $\delta{\bf n}$ is
\begin{equation}
\label{variaWZ}
\delta S_{WZ} = s\int_0^T dt~ \delta{\bf n}\cdot({\bf n}
\times \partial_t{\bf n})
\end{equation}

Next, we add interactions between spins on a cubic lattice $\Gamma'$, with
nearest neighbor vectors $(\hat{\bf x},\hat{\bf y},\hat{\bf z})$.
Anticipating short range antiferromagnetic order, we scale the unit cell by
a factor of 2 in every direction, thus obtaining a bigger lattice $\Gamma$
whose points are labeled ${\bf r}$ (unprimed) and whose unit cell contains
8 sites of $\Gamma'$:
\begin{equation}
{\bf r}' = \alpha\hat{\bf x}+\beta\hat{\bf y}+\gamma\hat{\bf z}
\end{equation}
wherein each of $\alpha$, $\beta$ and $\gamma$ runs from 0 to 1.
The unit cell of $\Gamma$ contains 8 spins which may be described by 8
differents vector fields, as follows:
\begin{mathletters}
\label{auxFields}
\begin{eqnarray}
&&{\bf S}({\bf r}') = s(-1)^{\alpha+\beta+\gamma}{\bf n}({\bf r}')\\
&&{\bf n}({\bf r}') = {\bf m}({\bf r}) + a\mathop{{\sum}'}_{i,j,k}
(-1)^{i\alpha+j\beta+k\gamma}{\bf l}_{ijk}({\bf r})
\end{eqnarray}
\end{mathletters}
Again each of $\alpha,\beta,\gamma$ and $i,j,k$ runs from 0 to 1;
we have introduced 7 deviation fields ${\bf l}_{ijk}$, the primed
sum meaning that the term $i=j=k=0$ is omitted, the latter being rather
represented by the slowly varying field ${\bf m}$. We have included a factor
of $a$ in the definition of ${\bf l}_{ijk}$ in order to stress that deviations
from short range antiferromagnetic order are assumed to be small.

Now we must express the kinetic term and the Heisenberg Hamiltonian in terms
of these new fields, at lowest non-trivial order in $a$, and then integrate
out the deviation fields ${\bf l}_{ijk}$ to obtain an effective continuum
action for ${\bf m}$. Let us begin with the kinetic term $K$, which is obtained
by summing over spins the action (\ref{actionWZ}). Assuming that ${\bf n}$ is
slowly varying, we use Eq. (\ref{variaWZ}) to write
\begin{eqnarray}
K =  s\sum_{{\bf r}\in\Gamma}\sum_{\alpha,\beta}&&(-1)^{\alpha+\beta}
\delta_{z}{\bf n}({\bf r}')\cdot({\bf n}({\bf r}')
\times \partial_t{\bf n}({\bf r}')) \nonumber\\
&&({\bf r}' = {\bf r} + \alpha\hat{\bf x}+ \beta\hat{\bf y})
\end{eqnarray}
where we defined the difference
\begin{equation}
\delta_z{\bf n}({\bf r}') =
{\bf n}({\bf r}'+\hat{\bf z})-{\bf n}({\bf r}')
\end{equation}
and similarly for $\delta_x$ and $\delta_y$.
The definitions (\ref{auxFields}) imply
\begin{equation}
\label{diffA}
\delta_z{\bf n}({\bf r}+\alpha\hat{\bf x}+\beta\hat{\bf y})
= -2a\sum_{i,j}(-1)^{i\alpha+j\beta}{\bf l}_{ij1}({\bf r})
\end{equation}
Applying the identity
\begin{equation}
\sum_{\alpha=0,1} (-1)^{\alpha(i+j)} = 2\delta_{ij}
\end{equation}
we may write the kinetic term as
\begin{equation}
K = 2^d as\sum_{{\bf r}\in\Gamma} {\bf l}_{111}({\bf r})\cdot
({\bf m}({\bf r})\times \partial_t{\bf m}({\bf r}))
\end{equation}
where $d$ is the dimension of space, which we keep variable even though our
notation is three-dimensional, in order to make this derivation more general.
In the continuum, this becomes
\begin{equation}
K = a^{1-d}s\int d{\bf r}~{\bf l}_{111}\cdot
({\bf m}\times \partial_t{\bf m})
\end{equation}

The Heisenberg Hamiltonian may be expressed as $H = V_x + V_y + V_z$ where, for
instance,
\begin{equation}
V_z = -s^2 J_z\sum_{{\bf r}'\in\Gamma'}{\bf n}({\bf r}')\cdot
{\bf n}({\bf r}'+\hat{\bf z})
\end{equation}
Up to an irrelevant constant, this is equal to
\begin{equation}
{\textstyle\frac12} s^2 J_z\sum_{{\bf r}'\in\Gamma'}
(\delta_z{\bf n}({\bf r}'))^2
\end{equation}
We have similar expressions for $V_{x,y}$ wherein $J_z$ and
$\delta_z$ are replaced by $J_{x,y}$ and $\delta_{x,y}$. We then use Eq.
(\ref{diffA}), along with
\begin{equation}
\label{diffB}
\delta_z{\bf n}({\bf r}+\alpha\hat{\bf x}+\beta\hat{\bf y}
+\hat{\bf z})  = 2a \partial_z{\bf m}({\bf r})
+2a\sum_{i,j}(-1)^{i\alpha+j\beta}{\bf l}_{ij1}({\bf r})
\end{equation}
to find
\begin{equation}
V_z =2^{d}a^2s^2J_z\sum_{{\bf r}\in\Gamma}\bigg\{
2\sum_{ij}{\bf l}_{ij1}^2 + 2{\bf l}_{001}\cdot\partial_z{\bf m} +
(\partial_z{\bf m})^2 \bigg\}
\end{equation}
along with similar expressions for $V_x$ and $V_y$.

In the continuum, we may therefore write the following Lagrangian density:
\begin{eqnarray}
{\cal L} &=& a^{1-d}s {\bf l}_{111}\cdot({\bf m}\times \partial_t{\bf m})
\nonumber\\
&&\quad - a^{-d} s^2 J_x\bigg\{2\sum_{i,j}{\bf l}_{1ij}^2 +
2{\bf l}_{100}\cdot\partial_x{\bf m}+ (\partial_x{\bf m})^2 \bigg\}\nonumber\\
&&\quad - a^{-d} s^2 J_y\bigg\{2\sum_{i,j}{\bf l}_{i1j}^2 +
2{\bf l}_{010}\cdot\partial_y{\bf m}+ (\partial_y{\bf m})^2 \bigg\}\nonumber\\
&&\quad - a^{-d} s^2 J_z\bigg\{2\sum_{i,j}{\bf l}_{ij1}^2 +
2{\bf l}_{001}\cdot\partial_z{\bf m}+ (\partial_z{\bf m})^2 \bigg\}
\end{eqnarray}

We now proceed to the functional integration of the fields ${\bf l}_{ijk}$.
Since the action is quadratic in these fields, this amounts to substituting
in the Lagrangian the expression of these fields obtained from the equations
of motion. The latter are
\begin{mathletters}
\label{eqmouv}
\begin{eqnarray}
&&{\bf l}_{111} =
{1\over 4as}{1\over J_x+J_y+J_z}{\bf m}\times \partial_t{\bf m}\\
&&{\bf l}_{100} = -{\textstyle\frac12} \partial_x{\bf m}\qquad
{\bf l}_{010} = -{\textstyle\frac12} \partial_y{\bf m}\qquad
{\bf l}_{001} = -{\textstyle\frac12} \partial_z{\bf m} \\
&&{\bf l}_{ijk} = 0\qquad{\rm otherwise}
\end{eqnarray}
\end{mathletters}
Substitution of these equations into the above Lagrangian density yields
exactly Eqs. (\ref{nonlinearS}) and (\ref{GandV}), with $J_z=J$,
$J_x=J_y=RJ$ and $d=3$.
Note that in the isotropic case ($J_n=J$), the Lagrangian obtained
in dimension $d$ is
\begin{equation}
\label{nonlinearSb}
{\cal L} = {1\over 2g}\left\{
{1\over v} (\partial_t{\bf m})^2 -
v\sum_i (\partial_i{\bf m})^2\right\}
\end{equation}
with
\begin{equation}
v = 2Jas\sqrt{d}\qquad
g = {2a^{d-1}\over s}\sqrt{d}
\end{equation}

The above derivation is valid for any dimension greater than 1. In dimension
1 a complication occurs since there is only one deviation field ${\bf l}$,
whose equation of motion is instead
\begin{equation}
{\bf l} = -{\textstyle\frac12} \partial_z{\bf m} +
{1\over 4aJs}{\bf m}\times \partial_t{\bf m}\end{equation}
Substitution into the Lagrangian yields, in addition to the action
(\ref{nonlinearSb}), a topological Hopf term, responsible for the difference
in behavior between integer and half-integer spin chains.
The above derivation shows clearly that no such term arises in dimensions
$d>1$.

So far we have eluded the question of constraints. The 8 fields ${\bf m}$ and
${\bf l}_{ijk}$ are related by 8 constraints coming from the relation
${\bf n}^2=1$. In the limit of small deviations ($a\to0$) these
constraints are \begin{equation}
{\bf m}^2 = 1\qquad\qquad {\bf m}\cdot{\bf l}_{ijk} = 0
\end{equation}
the second of these equations is compatible with the equations of motion
(\ref{eqmouv}), implying that the above substitution procedure was indeed
correct. The constraint ${\bf m}^2 = 1$ is part of the definition of the
nonlinear $\sigma$-model.

\section{Derivation of the gap equation}
The aim of this section is to find, starting from the $\sigma$-model
Lagrangian (\ref{nonlinearS}), an equation governing the temperature and
anisotropy dependence of the Haldane gap. Before embarking upon calculations,
let us simply state the results, obtained in the approximation where the cutoff
$\Lambda$ (proportional to the inverse lattice spacing $a^{-1}$) is much larger
than the temperature $T$ or the gap $\Delta$. Let $\Delta_0$ be the Haldane gap
at zero temperature in the one-dimensional limit ($R\to 0$), and let us
introduce the reduced variables and parameters
\begin{equation}
\label{redvar}
\delta = {\Delta\over\Delta_0}\qquad
t = {T\over\Delta_0}\qquad
r = {R\Lambda^2\over\Delta_0^2}
\end{equation}
Then the reduced gap $\delta$ obeys the following equation:
\begin{equation}
\label{gapEq}
(\delta^2+r)\ln(\delta^2+r) - \delta^2\ln\delta^2 - r +
8t^2\left[ G\left({\sqrt{\delta^2+r}\over t}\right)-
G\left({\delta\over t}\right)\right] = 0
\end{equation}
where we have defined the special function
\begin{equation}
\label{Gdef}
G(y) = {\textstyle\frac12}y^2 \int_1^\infty dz~\sqrt{z^2-1}\left[
\coth(yz/2)-1\right] = y\sum_{n=1}^\infty {1\over n}K_1(ny)
\end{equation}
($K_1$ is the modified Bessel function of order 1).
The remainder of this section will be devoted to the proof of Eq.
(\ref{gapEq}).

A large part of this section parallels the calculations presented in
Ref.\onlinecite{Senechal}. We start with the $\sigma$-model Lagrangian
(\ref{nonlinearS}) expressed in terms of a rescaled variable
$\varphi={\bf m}/\sqrt{g}$:
\begin{equation}
{\cal L} = {1\over 2}\left\{
(\partial_t\varphi)^2 -
(\partial_z\varphi)^2 -R (\partial_\bot\varphi)^2
-\sigma(\varphi^2-1/g) \right\}
\end{equation}
Here we have set the magnon speed $v$ to unity; it can be restored at the
end by dimensional analysis; we have also introduced a Lagrange multiplier
$\sigma$ whose r\^ole is to enforce the constraint $\varphi^2=1/g$.

The strategy used to compute the Haldane gap is to calculate at one loop
the effective potential $V(\sigma)$ for the Lagrange multiplier. This amounts
to integrating the field $\varphi$, assuming $\sigma$ to be constant.
We then look for a minimum in $V(\sigma)$. If such a minimum exists for
$\sigma\ne0$, then the position $\sigma$ of this minimum is the mass squared
of the triplet field $\varphi$, namely the square of the gap.

At one loop, the derivative $V'(\sigma)$ of the zero-temperature effective
potential in the Euclidian formalism is
\begin{equation}
V'(\sigma) = {1\over 2g} -3\int{dp_0\over (2\pi)}\int{d{\bf p}\over
(2\pi)^3} ~{1\over p_0^2 +p^2_z + R{\bf p}^2_\bot + \sigma}
\end{equation}
The gap equation is then simply $V'(\Delta^2)=0$, where $\Delta=\sqrt{\sigma}$.
The extension of this formula to a finite temperature $T$ is obtained by
replacing the integration over $p_0$ by a sum over Matsubara
frequencies $\omega_n = 2\pi nT$:
After summing over Matsubara frequencies, the gap equation becomes
($\beta=1/T$)
\begin{equation}
{1\over g} = 3\int{d{\bf p}\over(2\pi)^3}~
{1\over\sqrt{p^2_z + R{\bf p}^2_\bot + \Delta^2}}\coth{\beta\over 2}
\sqrt{p^2_z + R{\bf p}^2_\bot + \Delta^2}
\end{equation}
This expression is meaningless without the introduction of a momentum
cut-off $\Lambda$, which should be proportional to the inverse lattice
spacing $a^{-1}$. We will evaluate this integral by restricting it to
a cylinder of height $2\Lambda$ and radius $\Lambda$, parallel to the
${\bf p}_z$ axis in momentum space. Defining the integration variable $x=R{\bf
p}^2_\bot$, we have
\begin{equation}
{4\pi^2 R\over 3g} = \int_0^\Lambda dk~\int_0^{\Lambda^2R}dx~
{1\over\sqrt{k^2 + x + \Delta^2}}\coth{\beta\over 2}
\sqrt{k^2 + x + \Delta^2}
\end{equation}
The integral over $x$ can be done analytically:
\begin{equation}
\label{gapEqPa}
{\pi^2 R\over 3g} = T\int_0^\Lambda dk~\left\{
\ln\sinh{\beta\over2}\sqrt{k^2 +\Delta^2 + \Lambda^2R}-
\ln\sinh{\beta\over2}\sqrt{k^2 +\Delta^2}\right\}
\end{equation}
The remaining integral may also be done analytically, in the limit
$\Lambda\gg\Delta,T$: Let us introduce the notation
\begin{equation}
F(\alpha,\beta,\Lambda) =
\int_0^\Lambda dk~\ln\sinh{\beta\over2}\sqrt{k^2 +\alpha^2}
\end{equation}
After integrating by parts, this function becomes
\begin{equation}
F(\alpha,\beta,\Lambda) = \Lambda\ln\sinh{\beta\over2}\sqrt{\Lambda^2
+\alpha^2}
-{\beta\over2}\int_0^\Lambda{k^2 dk\over\sqrt{k^2 +\alpha^2}}
\coth{\beta\over2}\sqrt{k^2 +\alpha^2}
\end{equation}
The second term may be separated into a contribution at $T=0$:
\begin{equation}
\int_0^\Lambda{k^2 dk\over\sqrt{k^2 +\alpha^2}}
\approx {\textstyle\frac12}\Lambda^2 + {\textstyle\frac14}\alpha^2\left(
1 + \ln{\alpha^2\over 4\Lambda^2}\right)\qquad (\Lambda\to\infty)
\end{equation}
plus a finite temperature correction, finite as $\Lambda\to\infty$:
\begin{eqnarray}
&&\int_0^\infty{k^2 dk\over\sqrt{k^2 +\alpha^2}}\left\{
\coth{\beta\over2}\sqrt{k^2 +\alpha^2}-1\right\} \nonumber\\
&&\qquad = \alpha^2\int_1^\infty dz~\sqrt{z^2-1}\left[
\coth(\beta\alpha z/2)-1\right] \nonumber\\
&&\qquad = 2\alpha^2 G(\beta\alpha)
\end{eqnarray}
Summing up all contributions and expanding around $\Lambda=\infty$, we
obtain
\begin{equation}
F(\alpha,\beta,\Lambda\to\infty) =
{\textstyle\frac14}\beta\Lambda^2 - \Lambda\ln2 +{\textstyle\frac18}\beta
\alpha^2\left(1-\ln{\alpha^2\over 4\Lambda^2}\right) -
{1\over\beta}G(\beta\alpha)
\end{equation}
This result, inserted into Eq. (\ref{gapEqPa}), yields at last
\begin{eqnarray}
\label{gapEqB}
{8\pi^2 R\over 3g} &=&
(\Delta^2+R\Lambda^2)\left(1-\ln{\Delta^2+R\Lambda^2\over 4\Lambda^2}\right)
-\Delta^2\left(1-\ln{\Delta^2\over 4\Lambda^2}\right)\nonumber\\
&&\qquad -8T^2\left[ G\left({\sqrt{\Delta^2+R\Lambda^2}\over T}\right)-
G\left({\Delta\over T}\right)\right]
\end{eqnarray}
This is essentially the gap equation (\ref{gapEq}), except that the
reduced variables (\ref{redvar}) have not been introduced. To this end, let us
consider the one-dimensional case, i.e., let us take the limit $R\to0$. We must
use the fact that
\begin{equation}
{dG(y)\over dy} = -yH(y)\qquad\qquad H(y)=\sum_{n=1}^\infty K_0(ny)
\end{equation}
We then obtain the one-dimensional equation
\begin{equation}
{8\pi^2\over 3g\Lambda^2} = -\ln{\Delta^2\over 4\Lambda^2} + 4H(\Delta/T)
\end{equation}
Since $H(\infty) = 0$, the zero-temperature gap is
$\Delta_0=2\Lambda\exp-(4\pi^2/3g\Lambda^2) $. In terms of the reduced
variables (\ref{redvar}) the finite temperature gap equation in dimension 1
is then
$$ {\textstyle\frac12}\ln\delta = H(\delta/t) $$
This equation also appears in Ref.\onlinecite{Senechal}. Finally, upon
substitution of the reduced variables, the three-dimensional gap equation
(\ref{gapEqB}) takes the form (\ref{gapEq}).

\section{Discussion} %

In this section we illustrate the consequences of the gap equation
(\ref{gapEq}). That implicit equation may easily be solved numerically, given
the expression (\ref{Gdef}) for the function $G$ in terms of modified
Bessel functions. Fig. 1 shows the $t$ dependence
of $\delta$ for various values of the anisotropy parameter $r$.

At zero temperature, the gap equation reduces to
\begin{equation}
(\delta^2+r)\ln(\delta^2+r) - \delta^2\ln\delta^2 - r = 0
\end{equation}
This equation has the solution $\delta=1$ in the limit $r\to0$, and has a
nonzero solution for sufficiently small $r$. Fig. 2 illustrates the numerical
solution of $\delta$ vs $r$.
Taking the limit $\delta\to 0$ we see that the critical value of the
anisotropy parameter is $r_c=e$. This in turn corresponds to a critical
value $R_c$ of the coupling ratio $R=J_\bot/J$.

Beyond this value of $r$, a gap will appear only above some critical
temperature $t_c$ (the N\'eel temperature). If we take the limit $\delta\to0$
in Eq. (\ref{gapEq}), we obtain
\begin{equation}
\label{TcEq}
r\ln r - r +
8t_c^2\left[ G\left({\sqrt{r}\over t_c}\right)-\pi^2/6\right] = 0
\end{equation}
wherein we have substituted the special value $G(0)=\pi^2/6$.
This equation may be solved numerically for $t_c$ as a function of $r$,
and the result is illustrated on Fig. 3.

Let us now discuss the relation existing between the critical value $r_c$
and the microscopic parameters $J$ and $R=J_\bot/J$. The coupling $J$
resurfaces if the magnon speed $v$ is restored in the explicit value of the
gap by dimensional analysis. The expression for $\Delta_0$, the
zero-temperature gap in the $R\to0$ limit, is then
\begin{equation}
\Delta_0=2v\Lambda\exp-(4\pi^2/3g\Lambda^2)
= 4J\Lambda as\exp-\left[4\pi^2s/6(a\Lambda)^2\right]
\end{equation}
The ratio $\Delta_0/J$ can then be related to the cut-off prescription,
i.e., to the product $\Lambda a$. If we accept the ratio obtained from
numerical simulations\cite{Night}, namely $\Delta_0 = 0.41J$, we are led
by the above equation to the prescription $\Lambda a=2.1$. Notice however
that the ratio $\Delta_0/J$ is extremely dependent on the value of $\Lambda a$,
because of the exponential factor.

After restoration of $v$, the expression for the anisotropy parameter $r$ is
\begin{equation}
r = {\textstyle\frac14}R\exp-\left[8\pi^2s/6(a\Lambda)^2\right]
\end{equation}
The critical value $r_c=e$ then translates into a critial ratio
\begin{equation}
R_c = 4\exp\left[1 - 8\pi^2s/3(a\Lambda)^2\right]
\end{equation}
The choice $\Lambda a=2.1$ then leads to $R_c=.026$. This value, although
small, seems too high when compared to coupling ratios observed in actual
materials; for instance, the ratio $R$ has been estimated\cite{Buyers} to be
$\sim 0.02$ in CsNiCl$_3$, in which a finite N\'eel temperature is observed.
Of course, CsNiCl$_3$ and other compound of the ABX$_3$ type have a hexagonal
structure, not cubic.\cite{noteA}
It seems that trying to find the correct value of $R_c$ from a well-chosen
cut-off prescription is dangerous. However, definite predictions may be
obtained from a continuum theory, up to an overall scale in all but one
variables.

\acknowledgements  This work is supported by NSERC and by F.C.A.R. (le Fonds
pour la Formation de Chercheurs et l'Aide \`a la Recherche du Gouvernement du
Qu\'ebec).%
%

%
\begin{figure}
\caption{
Reduced Haldane gap $\delta$ as a function of the reduced temperature $t$
for various values of the anisotropy parameter $r$: 0.01, 1.5,
$e=2.718\ldots$ and $4$.
}
\end{figure}
\begin{figure}
\caption{
Reduced N\'eel temperature $t_c$ as a function of the anisotropy parameter
$r=R\Lambda^2/\Delta^2_0$.
}
\end{figure}
\begin{figure}
\caption{
Magnitude of the reduced Haldane gap $\delta=\Delta/\Delta_0$ at zero
temperature as a function of the anisotropy parameter
$r=R\Lambda^2/\Delta^2_0$.
The critial value of $r$ is $e$.
}
\end{figure}

\begin{references}
\bibitem[*]{noteA} The antiferromagnetic frustration in ABX$_3$ compounds
should lead to a long distance field theory different from the nonlinear
$\sigma$-model, with a different order parameter. Whether or not this should
have an influence on the temperature dependence of the gap is debatable.
\end{references}
\end{document}